\documentclass{webofc}
\usepackage[varg]{txfonts}   
\usepackage{subcaption}
\usepackage[font=small,labelfont=bf,textfont=normal,singlelinecheck=false]{caption}
\usepackage{natbib}
\usepackage[dvipsnames]{xcolor}
\usepackage{array}
\usepackage{booktabs}
\usepackage{hyperref}
\usepackage{makecell}
\usepackage{wrapfig, blindtext}
\usepackage{graphicx}
\usepackage{minipage}
\usepackage{amssymb}
\usepackage{enumerate}
\usepackage{pifont}
\newcommand{\cmark}{\ding{51}}
\newcommand{\xmark}{\ding{55}}
\usepackage{inputenc}
\usepackage{titlesec}
\titlespacing\section{0pt}{12pt plus 4pt minus 2pt}{6pt plus 2pt minus 2pt}
\titlespacing\subsection{0pt}{12pt plus 4pt minus 2pt}{0pt plus 2pt minus 2pt}
\titlespacing\subsubsection{0pt}{12pt plus 4pt minus 2pt}{0pt plus 2pt minus 2pt}
\newcommand{\ok}{\textbf{\textcolor{ForestGreen}{OK}}}
\newcommand{\warning}{\textbf{\textcolor{BurntOrange}{WARNING}}}
\newcommand{\critical}{\textbf{\textcolor{BrickRed}{CRITICAL}}}

\begin{document}
\title{Benchmarking NetBASILISK: a Network Security Project for Science}
\author{
\firstname{Jem} \lastname{Guhit} \inst{1},
\firstname{Edward} \lastname{Colone} \inst{2}, 
\firstname{Shawn} \lastname{McKee} \inst{1},
\firstname{Kris} \lastname{Steinhoff} \inst{2,3},
\firstname{Katarina} \lastname{Thomas} \inst{2}
}

\institute{Physics Department, University of Michigan, Ann Arbor, MI, USA
\and
Information and Technology Services, University of Michigan, Ann Arbor, MI, USA
\and 
School of Information, University of Michigan, Ann Arbor, MI, USA
}

\abstract{
Infrastructures supporting distributed scientific collaborations must address competing goals in both providing high performance access to resources while simultaneously securing the infrastructure against security threats. The NetBASILISK project is attempting to improve the security of such infrastructures while not adversely impacting their performance.  This paper will present our work to create a benchmark and monitoring infrastructure that allows us to test for any degradation in transferring data into a NetBASILISK protected site.
}
\maketitle

\section{Introduction}
Distributed computing infrastructures face significant challenges in effectively supporting scientists in moving, accessing, analyzing and transforming data to produce new scientific insights. This is complicated by continuous attacks that all systems connected to the Internet experience\cite{JANGJACCARD2014973}\cite{SigCyberIncidents}. The challenge is to secure the infrastructure without compromising the performance and usability of the infrastructure.

The Energy Sciences network (ESnet)\cite{ESnet} created a new paradigm in 2010 to help address this situation: the Science DMZ \cite{ScienceDMZ}.   A network DMZ, or demilitarized zone, is a physical or logical sub-network that contains and exposes an organization's external-facing services to an un-trusted network such as the Internet. The DMZ functions as a small, isolated network positioned between the Internet and the private network \cite{DMZ}.  This paradigm allowed data transfer nodes to bypass firewalls and security devices which would otherwise interfere with science data-flows, adversely impacting scientists in their work.  The intent is to minimize the number of devices attached to the Science DMZ and carefully monitor and configure them to provide both performance and security.

The NetBASILISK \cite{NetBASILISK} (NETwork Border At Scale Integrating and Leveraging Individual Security Components) project, led by researchers and network engineers at the University of Michigan, seeks to augment and expand the Science DMZ concept. The goal is to allow institutions to maintain both security and capability for all their users by prototyping and deploying a network border security solution capable of supporting unfettered network traffic at 4x100 Gbps using a mix of commercial offerings and locally-developed middleware.

In this paper we will describe the work we have done to evaluate the effectiveness of the NetBASILISK infrastructure and verify that it does not adversely impact the ability of the ATLAS scientists at the University of Michigan to move data from outside our institution.

\section{Rationale and Project Components} \label{Sec2}
A network security infrastructure has the potential to impact data entering or leaving an institution.  The focus of the work described in this paper is evaluating the impact of NetBASILISK on data coming into the University of Michigan.

To understand if NetBASILISK impacts the transfer of external data to the university, we need to:
\begin{itemize}
    \setlength\itemsep{-0.3em}
    \item Determine normal transfer behavior in the absence of NetBASILISK (get a baseline).
    \item Monitor relevant factors \textbf{ other} than NetBASILISK that might impact transfers.
    \item Create a benchmark test which can be run on-demand to determine transfer rates.
\end{itemize}

Using this set of information and tools, we can evaluate if NetBASILISK (or changes in NetBASILISK) are adversely impacting our ability to transfer data to the University of Michigan. The two capabilities we need to evaluate NetBASILISK's potential impact are a {\bf benchmark application}, which can be used both to derive a baseline and make on-demand evaluation of the current transfer performance, and an {\bf environmental monitoring application} which can gather and track relevant metrics that might be related to the observed transfer performance. In the following section we will describe the development of the benchmark and, later, the environmental monitoring application development.

\section{Development of the Benchmark}\label{BenchmarkDev}
 
The NetBASILISK benchmark application aims to build a stable transfer file system that would derive a baseline for normal transfer behavior and evaluate transfer performance. There are two main components to the benchmark application: \textbf{data} and \textbf{file transfer tool}.
\subsection{Data}
Since the University of Michigan Physics department is a collaborator of the ATLAS Experiment \cite{ATLAS}, the data would be coming from a repository containing samples of ATLAS datasets located at Brookhaven National Laboratory (BNL). BNL is an ideal choice for an external data source because it is one of the primary collaborators for data operations, it has a well-defined network for data transfers coming into the University of Michigan, and it already has a data repository intended to be accessible for long-term that could be used for the benchmark test.
\subsection{Transfer Tool}
After securing an external data source, a software framework tool for grid computing is used to access the ATLAS data repository and transfer files from BNL to the University of Michigan. Various software tools exist to handle file transfers. We decided to choose from the five common transfer tools used in the High Energy Physics Computing toolkit: File Transfer Service (FTS) \cite{FTS}, Storage Resource Management (SRM) \cite{SRM}, Grid File Access Library (GFAL) \cite{GFAL}, Globus GridFTP \cite{Globus}, and XRootD \cite{XRootD}.

To ensure that the benchmark test is reliable and consistent after successive test runs, we imposed a criteria that the transfer tool has to satisfy, such as: 
\begin{enumerate}[a.]
    \setlength\itemsep{-0.3em}
    \item Stable Transfer Speed
    \item Consistency
    \item Checksum Verification
    \item Information about the destination storage nodes
\end{enumerate}
\subsubsection{Transfer Tool Evaluation}
The transfer tool's capabilities are tested by running multiple data transfers using a subset from the ATLAS data repository, specifically transferring the same ten datasets ten times. Originally we planned to run the test only once, however, two transfer tools satisfied the criteria shown in Table \ref{tab:table1}, thus another round of the test had to be done. At the end, Transfer Tool Evaluation evolved to encompass  two test rounds: the first focused on whether the storage clients satisfy the criteria, second  tested the variability and consistency of the storage clients that passed the initial criteria.
\begin{table}[htbp]
\centering
\begin{tabular}{|c|c|c|c|c|} 
\toprule
\textbf{Storage Client} &  \multicolumn{4}{c|}{\textbf{Criteria}} \\
\midrule
\hline
{}  & \makecell{a. Transfer \\ Speed} & b. Consistency  & \makecell{c. Checksum \\ Verification}  & \makecell{d. Storage \\ Node}\\
\hline
FTS   &  \xmark & \xmark   & \cmark  & \xmark\\
\hline
SRM   &  \xmark & \cmark  & \cmark & \xmark\\
\hline
GFAL   &  \cmark &  \cmark   & \cmark  & \xmark\\
\hline
Globus   &  \cmark &  \cmark   & \cmark  & \cmark \\
\hline
XRootD   &  \cmark &  \cmark   & \cmark  & \cmark\\
\hline
\bottomrule
\end{tabular}
\caption{Results of the first round of the storage client test. The table shows the requirements satisfied by each storage client.}
\label{tab:table1}
\vspace{-1.5em}
\end{table}

Table \ref{tab:table1} summarizes the first test round: it shows how different transfer tools satisfy the established criteria . From the first round, three storage clients were ruled out: FTS, SRM, and GFAL. FTS did not satisfy criteria (a), (b), and (d): its non-trivial way to extract the transfer rate and destination storage node information from the log file failed criteria (a) and (d), FTS' own system for scheduling data transfers would limit our flexibility to control data transfers thus it failed criterion (b). SRM did not satisfy criteria (a) and (d): for (a), the protocol revealed a huge variability in transfer rates, while for criterion (d) it had difficulties in extracting the destination storage node information from the log files. GFAL did not satisfy criterion (d), because it did not provide information about the destination storage node. Globus and XRootD both satisfied all the criteria and went to the next round of testing.

\begin{figure}[th]
\centering
\includegraphics[width=1\textwidth]{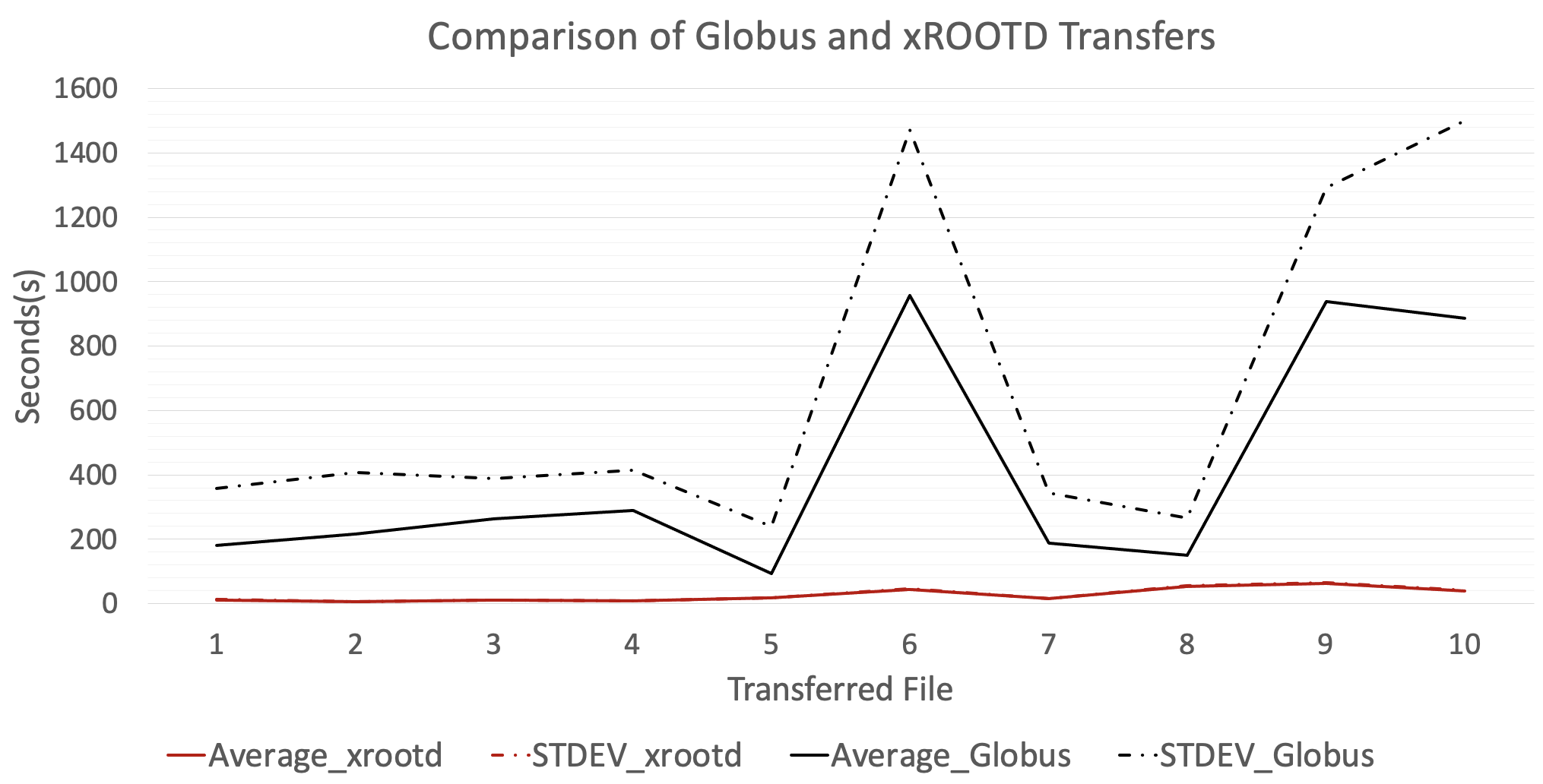}
\caption{The tests transferred the same ten files ten times for each tool. The STDEV and Average for the xROOTD case are almost overlapping. The figure shows that Globus has more fluctuation and instability compared to xROOTD. The total time for 10 tests for Globus is 11.5 hours while total time for 10 tests in xROOTD is 3.79 hours. }
\label{fig-1}       
\vspace*{-15pt}
\end{figure}

The second test iteration, although virtually same as the round one, focused  on finding out which storage client could provide the most stable file transfer performance over a period of time. In addition, information such as the average and standard deviation for the transfer rates was gathered for comparison. Figure \ref{fig-1} shows that after the second round of test between Globus and XRootD, the average and standard deviation of the former had shown higher transfer rate variability. This determined that XRootD protocol is better suited to use as the official transfer tool for the benchmark application.

\section{Environmental Monitoring Application} \label{Environment}
Having a benchmark to evaluate transfer performance is critical for our task but it is equally important to understand the environment in which the benchmark is run. A poor transfer result could be because NetBASILISK has adversely impacted the transfer {\bf OR} it may be due to a number of other possibilities:
\begin{itemize}
    \setlength\itemsep{-0.3em}
    \item The network between the benchmark source data and  destination may be congested at the source or destination local area network, or in the wide area network.
    \item The source data servers might be overloaded or have other resource issues.
    \item The destination data servers might be overloaded or have other resource issues.
\end{itemize}

To determine the cause of a poor benchmark result we have to gather sufficient information about the environmental conditions under which the benchmark was executed. Thus we need information on the end systems and the network in between, including the Brookhaven National Laboratory (BNL) storage systems (source of the benchmark data); the network path between BNL and the ATLAS Great Lakes Tier-2 \cite{AGLT2} cluster at the University of Michigan; and the storage nodes at AGLT2 (destination for the benchmark data). 
\subsection{Network Monitoring}
To monitor the network status, we needed to identify the existing network monitoring mechanisms along our network path. For the environment monitoring script, the path segments were defined as the source network at BNL, the WAN between our sites and the destination network at AGLT2. Although we considered measuring the round-trip time (RTT), we found that it was not as relevant for our use-case.  

\label{NetBNL}
{\bf Network at Source:} While network monitoring internal to BNL exists,  we didn't have easy access to those systems. The internal networking at BNL is well provisioned and we are assuming that the BNL network will not limit our benchmark transfers.  As we continue to evolve the environment monitoring, we intend to include explicit BNL network monitoring.

{\bf Wide-area Network:} \label{WAN} The wide area network path between BNL and AGLT2 has two distinct components:  1) a path from BNL to Chicago on ESnet, and 2) a path from ESnet's location in Chicago to AGLT2. We will discuss each components below. 
\begin{figure}[th]
\centering
\includegraphics[width=0.95\textwidth]{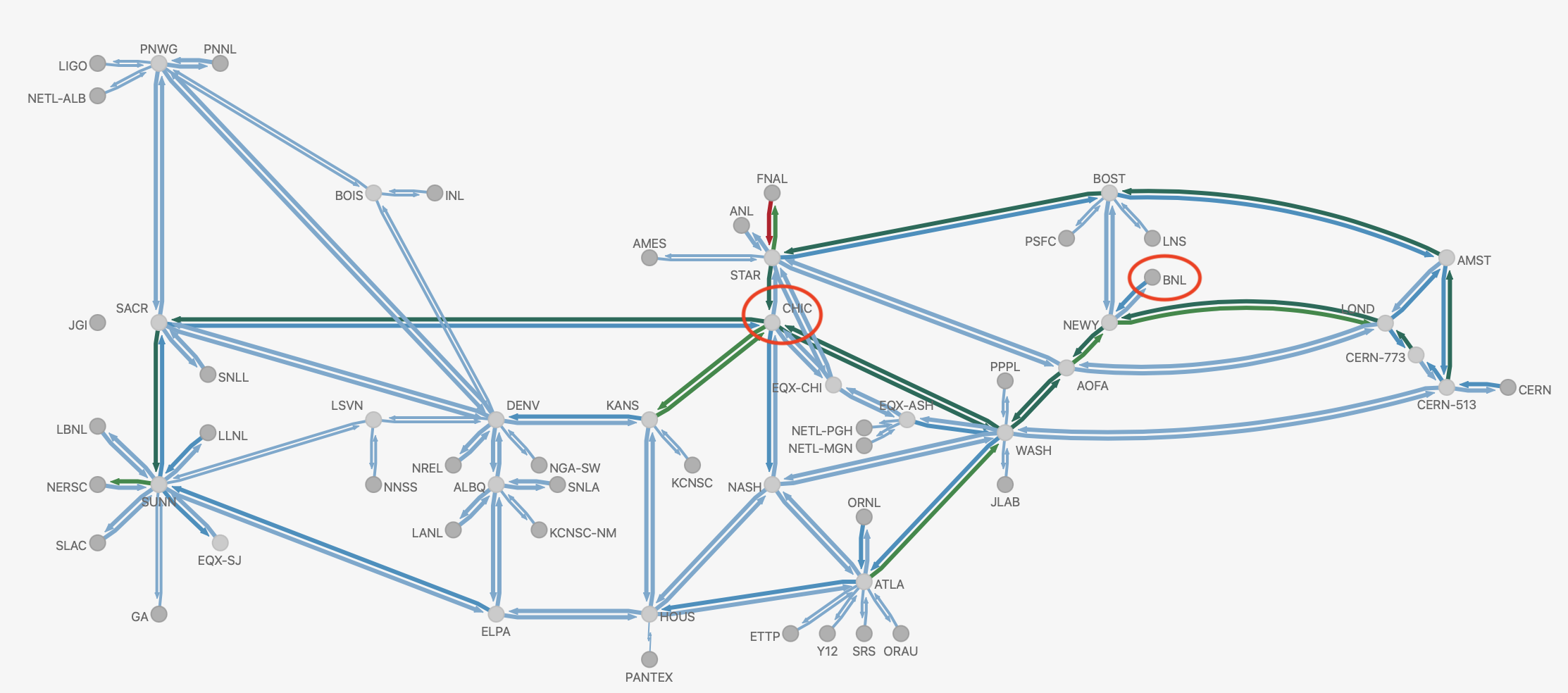}
\caption{The ESnet network monitored on their portal. Shown are multiple paths from BNL to CHIC (Chicago), emphasized in red circles, where our benchmark traffic would exit ESnet to connect to AGLT2. ESnet uses internal traffic engineering which precludes us from knowing which of the many possible physical paths between BNL and CHIC are followed for a given transfer.} 
\label{fig-2}  
\vspace*{-15pt}
\end{figure} 
For the 1) BNL to Chicago component we encountered some challenges in monitoring the network path.  ESnet, which handles the part of the path, provides publicly accessible monitoring information\cite{MyESnet}. Figure \ref{fig-2} shows the ESnet traffic monitoring, including the possible paths between BNL to Chicago (CHIC) from the \hyperlink{https://my.es.net}{my.es.net} website. Unfortunately, because ESnet uses MPLS\cite{MPLS} to control traffic between their ingress and egress points, we don't have a way to identify {\bf which} interfaces and paths our benchmark traffic traversed.  In our current environment monitoring, we assume the ESnet portion of the path will not be congested and not contribute to a poor benchmark result due to their traffic engineering.  We have been in contact with the ESnet developers and will try to include explicit monitoring in a future update. 
\vspace{-2pt}
\begin{wrapfigure}[27]{l}{0.5\textwidth} 
\vspace{-15pt}
  \begin{center}
    \includegraphics[width=0.5\textwidth]{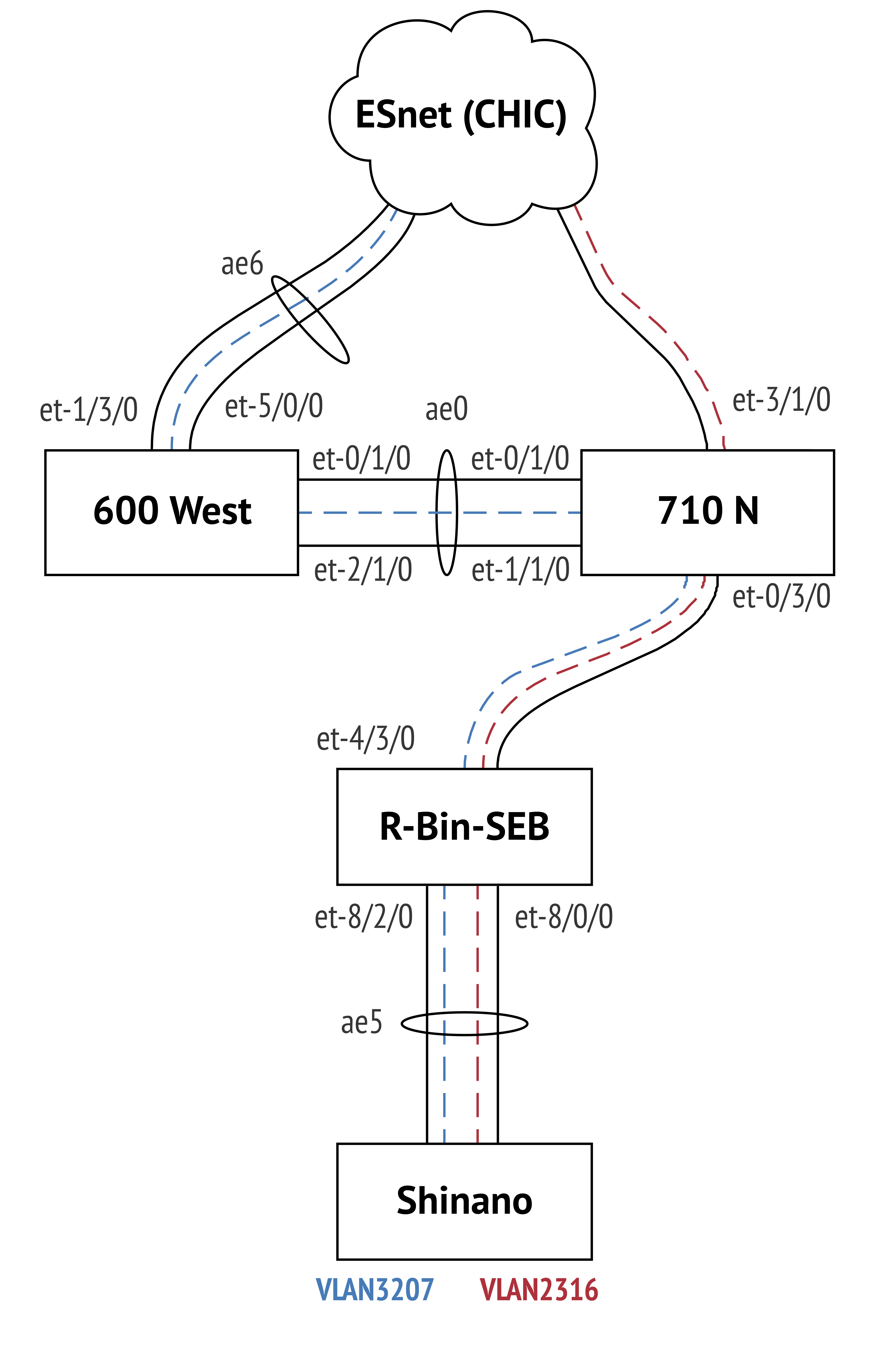}
  \end{center}
  \vspace{-15pt}
  \caption{Network Topology between Shinano (AGLT2) and ESnet (CHIC) showing the internal and external routers, ports, port-chanels, and VLAN line for each server.}
  \label{fig-3}  
\end{wrapfigure} 

The second part of the WAN path, Chicago to AGLT2, has more useful monitoring available. We monitor and verify that all the interface along our path are not congested. In Figure \ref{fig-3} we can show the routers and interfaces involved, starting from the ESnet presence in Chicago (at the top of the figure) and connecting all the way to the AGLT2 border router (named "Shinano") at the bottom.
The environment metrics of the Shinano Router (and AGLT2 LAN) are available from the CheckMK interface \cite{checkmk} and is discussed in more detail in Section \ref{CheckMK}. R-Bin-SEB is an internal router for the University of Michigan and monitored by the CA Performance Center.  We gather the following metrics using a python script developed to access the Application Programming Interface (API) of the CA Performance Center \cite{caperfscript}:
\begin{itemize} 
    \setlength\itemsep{-0.3em}
    \item Bits In/Out: converted to GBs In/Out
    \item Bits/s In/Out: converted to GB/s In/Out 
    \item Utilization In/Out: \% of capacity
    \item Error In/Out: number of packets that caused an error (e.g failed checksum)
\end{itemize} 
Buffer Overflow Discards, where packets are dropped when the buffer on either the incoming or outgoing port is full, is another metric being considered for R-Bin-SEB and will be included if accessible. The routers in Chicago (710 N / 600 West) have metrics stored in Grafana \cite{Grafana}.  We gather the following metrics from a python script developed to interact with Grafana \cite{grafanascript}:  
\begin{itemize}
     \setlength\itemsep{-0.3em}
    \item Input: measured in GBps
    \item Output: measured in GBps
\end{itemize}
\subsection{End-system Monitoring} \label{CheckMK}
In addition to the network we need to understand the operating conditions of the end-systems at the source and destination sites.

{\bf BNL End-System:} As mentioned in Section \ref{NetBNL}, even though a monitoring interface exists to capture metrics from BNL, accessing the system is currently non-trivial. In the future update, we plan to include the environment metrics for BNL and the script to access the API interface.

{\bf AGLT2 End-System:} As mentioned in Section \ref{WAN}, the IT Infrastructure monitoring software for the AGLT2 Cluster System is Check MK \cite{checkmk}. All transferred files sent to the AGLT2 Cluster are randomly stored in one of the dCache \cite{dCache} servers. Check MK contains a variety of network metadata of each dCache server and the metrics of interest for the benchmark script would be the the following: 
\begin{itemize}
    \setlength\itemsep{-0.3em}
    \item {\bf CPU Load}: number of active processes in the system.
    \item {\bf CPU Utilization}: estimates system performance in percentage.
    \item {\bf Memory}: variables of interest are MemFree and MemAvailable measured in GB.
    \item {\bf Disk Input Output (IO)}: contains the read and write operations of the physical disk and is usually measured in GBps. The variable of interest for us is the Disk Utilization measured in percentages.
\end{itemize} 

A python script was developed to gather these metrics from the Application Programming Interface (API) of the CheckMK Interface\cite{checkmkscript}.

The environment metrics described in this section are important components that would allow us to monitor the network status along the transfer path. The following section will describe how the benchmark application and environment monitoring application come together as a whole framework. 
\section{Framework Implementation} \label{BenchmarkImpl}
\begin{figure}[th]
\centering
\includegraphics[width=1.0\textwidth]{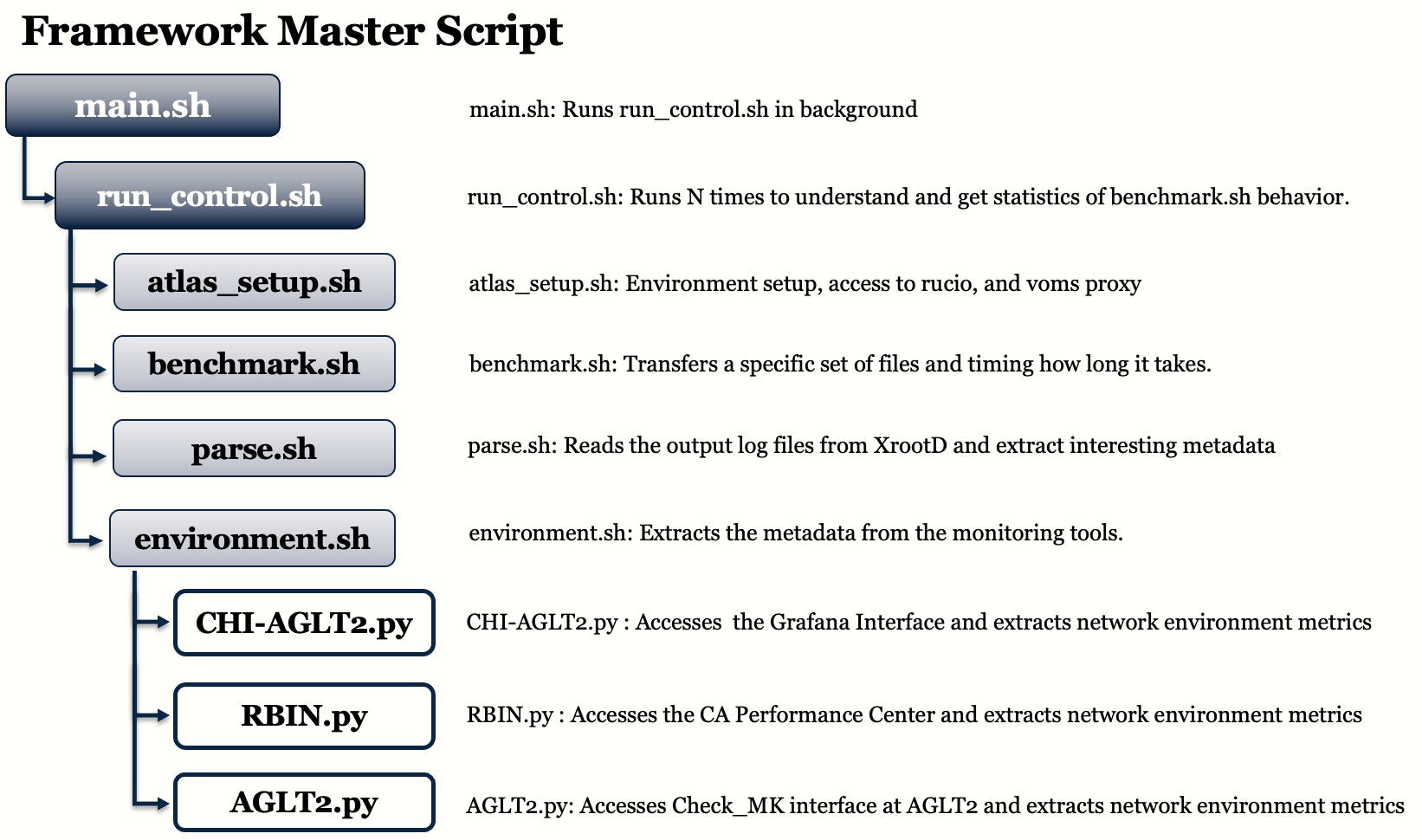}
\caption{First version of the framework master script where it runs the benchmark application followed by the environment monitoring application.} 
\vspace{-2.5em}
\label{fig-4}  
\end{figure}

The benchmark application discussed in Section \ref{BenchmarkDev} and the environment monitoring application in Section \ref{Environment} together are the backbone of the framework composed of shell and python scripts shown in Figure \ref{fig-4}.
It shows the first version of the framework script where it runs the benchmark application followed by the environment monitoring application.

The  \texttt{main.sh} script holds the framework together and runs the scripts in background to ensure running processes do not get aborted automatically when the SSH connection is lost. It contains a sub-module \texttt{run$\_$control.sh}, which runs four sub-modules. The \texttt{atlas$\_$setup.sh} sets up the XRootD software package. The \texttt{benchmark.sh} script is the beginning of the benchmark application component and handles the file transfers between BNL and AGLT2. This script also records the \texttt{Start} and \texttt{End} times of each benchmark test to be used for the \texttt{environment.sh} module. Each benchmark test produces a log file which contains important information about the transfer, more on Section \ref{TransferFileSys} which are extracted in \texttt{parse.sh} and organizes the output metadata into json files. This finishes the benchmark application portion and we move on to the environment monitoring application component.

As mentioned earlier, the \texttt{Start} and \texttt{End} times recorded from \texttt{benchmark.sh} are used as input variables for the three sub-modules of the \texttt{environment.sh}. We initially implemented that the environment monitoring application captures the network metrics in the same time-range the benchmark test was ran. The environment monitoring component has three independent scripts for the AGLT2 End-System (\texttt{AGLT2.py}) and Wide Area Network path for Chicago to AGLT2 (\texttt{CHI-AGLT2.py}, \texttt{RBIN.py}). These scripts have a similar structure of capturing network metrics but had to be scripted independently because of different API syntax of the network interfaces. 

The first version included the environment monitoring application in the framework script. However, we wanted to track the variability of the environment monitoring application in the long-term and not only when the benchmark application was ran. Therefore, we decided to detach the environment monitoring script from the benchmark application script and make it its own continuous operating service.
\subsection{Benchmark Application Metrics} \label{TransferFileSys}
As described above, a bash script \texttt{parse.sh} extracts the following metadata from the benchmark file transfers: 
\begin{itemize}
    \setlength\itemsep{-0.3em}
    \item {\bf Bandwidth}: The average rate of the benchmark transfers measured in MB/s.
    \item {\bf Transfer Time}: Total time to transfer all files measured in seconds. 
    \item {\bf Destination Host Frequency}: Count of files sent to each destination host. 
    \item {\bf File Size}: The size of each file transferred measured in GB. Summing gives us the total.
    \item {\bf Transfer Speed}: The transfer speed of each file measured in MB/s. 
\end{itemize}

These benchmark metrics are stored in a json file on the execution host and sent to Humio \cite{humio} for plotting and analysis. The benchmark results correspond to a benchmark score determining whether the average transfer speed of the test is \ok, \warning, or \critical. 

\subsection{Environment Monitoring Application Metrics}
As described above, a bash script \texttt{environment.sh} runs the three sub-modules: \texttt{CHI-AGLT2.py, RBIN.py, and AGLT2.py} to extract the environment metadata discussed in Section \ref{Environment}. \texttt{CHI-AGLT2.py} extracts the metadata from routers in Chicago (710N /600 W) stored in the Grafana interface and saves it in json format. Similarly \texttt{RBIN.py} extracts metadata from the R-Bin-SEB router stored in the CA Performance Center. A pre-processing script \cite{preprocessing} takes the json outputs and calculates important statistics (mean, standard deviation, min, and max). These results correspond to a benchmark score determining whether the network activity in the Chicago and AGLT2 path is \ok, \warning, or \critical. 

As for the AGLT2 End-System, \texttt{AGLT2.py} extracts the benchmark metadata defined in Section \ref{TransferFileSys} for each dCache server from the Check MK interface and saves it in json format. A pre-processing script \cite{preprocessing} takes the json output and calculates the statistics (mean, standard deviation, min, and max) for each dCache server. Since the transferred files are randomly assigned to the dCache servers, the \textbf{destination of host frequency} variable defined in Section \ref{TransferFileSys} is used to normalize and re-weight the statistics. These results correspond to the benchmark score determining whether the AGLT2 network activity at a specific time frame is \ok, \warning, or \critical. 
\section{Interpreting Results}
This section describes how we interpret the results coming from the benchmark application and environment monitoring application components. 
\subsection{Framework Score} 
The framework score is based on the results of the \textbf{(1)Benchmark Application} which provides statistics from the file transfers from  BNL to AGLT2. Then \textbf{(2) Environment Monitoring Application} which queries network metadata from the \textbf{(2)a Source}, \textbf{(2)b Network In Between}, and \textbf{(2)c Destination}. 

Three threshold levels are defined for each component: \ok, \warning, and \critical. \ok\ means fast transfer time in the context of (1) and a congestion free network in the context of (2)a - (2)c. \warning\ means medium transfer time in the context of (1) and a medium-congested network in the context of (2)a - (2)c. \critical\ means slow transfer time in the context of (1) and a congested network in the context of (2)a - (2)c. The threshold level for (2) is determined by the worst result of (2)a - 2(c).
\subsection{Framework Score Example}
How is the Benchmark Score interpreted to provide information about NetBASILISK? This is best explained by two cases presented in Table \ref{tab:table2}. Case 1 shows that the cause of a \critical\ transfer time might be due to NetBASILISK. The Testing Environment is designated as \ok, meaning the environment metrics indicate a congestion free network. However, the transfer time is slow which means that NetBASILISK might have played a role in this performance degradation. Case 2 indicates that even if the time transfer is \critical, the slow transfer time might not be caused by NetBASILISK as the Testing Environment is \critical\ also: the environment metrics indicate a congested network which might have been a cause for the slow transfer time. 
\begin{table}[htbp]
\centering
\begin{tabular}{|c|c|c|} 
\toprule
\textbf{Framework Components} &  \textbf{Case 1} &  \textbf{Case 2}   \\
\midrule
\hline
Source   &  \ok & \critical   \\
\hline
Network In Between &  \ok & \warning  \\
\hline
Destination   &  \ok &  \ok   \\
\hline
\textbf{Testing Environment} & \ok & \critical \\
\hline 
{}   &  {} &  {}  \\
\hline
Ave Transfer Speed   &  \critical &  \critical  \\
\hline
\bottomrule
\end{tabular}
\caption{In Case 1, all environment metrics are OK but poor transfer results are observed, indicating a likely problem caused by NetBASILISK.  Case 2 shows an example where the Source is the likely cause of the poor transfer result.}
\label{tab:table2}
\end{table} 
\section{perfSONAR Testing}
In order to correlate data transfer benchmarks with prevailing network conditions, NetBASILISK uses perfSONAR \cite{ps} to run periodic latency and bandwidth tests between the relevant network locations. perfSONAR is an open-source network metric gathering toolkit that allows for ad-hoc and scheduled testing between perfSONAR testpoints. It uses existing, standard tools such as iperf3 and owamp to generate performance metrics.

The benefit of using perfSONAR is to generate network performance metrics which can be contrasted against application- and service-level metrics. In theory, if we observe poor service metrics, we can determine if the network performance has impacted the service by using perfSONAR to generate throughput and latency measurements around the time of the substandard performance. If the network metrics are within the established performance thresholds, we can infer that other infrastructure related to the service, such as CPU, disk, relative load, and other conditions not related to the network are the cause of the poor service performance. If the network performance metrics are below the threshold, we can assert that the network itself is partially or wholly the cause of poor service performance.

To gauge NetBASILISK’s ipact on performance, we will run scheduled metric gathering tests.  This testing activity will alternate between simulated science activities interspersed with perfSONAR throughput and latency measurements. As AGLT2 represents data ingress, we will run file transfers from BNL to the University of Michigan. We will also utilize publicly available perfSONAR nodes to test throughput and latency from BNL to the University of Michigan to gauge metrics before and after the file transfers. This scheduled activity is run with cron at AGLT2. The script that is executed is an Ansible playbook \cite{ansible-playbook} that alternates perfSONAR testing with file transfer test activities. The Ansible playbook has provisions for interacting with pScheduler, and the benchmark scripts.
\section{Conclusion and Future Work}
To relate test result data with other science drivers, the NetBASILISK project will explore the use of the Humio log platform \cite{humio} to aggregate test results from a variety of benchmark and perfSONAR tests. 

We plan to update our benchmark implementation to automatically send results to the Humio API to include those test results in the repository shared with other parts of the NetBASILISK project. This work will involve evaluating the best ways to represent the benchmark outcome data in a way that it can be effectively analyzed.

The perfSONAR project is transitioning to the Elastic Stack platform \cite{elasticstack} for data archiving and visualization. The perfSONAR team has published Logstash pipelines in order to facilitate data analysis of perfSONAR results \cite{ps-archiving-sandbox}. To this end we are standing up an Elastic Logstash server and using the perfSONAR pipelines. Humio is able to ingest data with the same methods as Elastic, so we are going to use Logstash to import the data to Humio.

The ultimate goal is to be able to combine similar benchmark and network test results into a single repository. This will allow analysis across the variations and could ultimately provide for more detailed monitoring and alerting.  We hope to use Humio's capabilities as a data visualizer to contrast science driver results with perfSONAR and other metrics.
\section{Acknowledgements}
\enlargethispage*{4mm}
We gratefully acknowledge the National Science Foundation which supported this work through NSF grant OAC-1925476.   In addition, we acknowledge contributions from the rest of the NetBASILISK team.
\newpage
\bibliography{bibliography}
\end{document}